\documentclass[a4paper,11pt]{article}
\usepackage{pos}

\title{Photo-production of axions and neutrinos in compact objects}

\author*[a]{Miguel Vanvlasselaer}

\affiliation[a]{Theoretische Natuurkunde and IIHE/ELEM, Vrije Universiteit Brussel,
\& The International Solvay Institutes, Pleinlaan 2, B-1050 Brussels, Belgium }

\emailAdd{miguel.vanvlasselaer@vub.be}

\abstract{Compact stellar objects like supernovae and neutron stars cool by the emission of light particles like neutrinos and axions from their very dense interiors.  In this article, we study in detail the photo-production of axions and neutrinos. We  point out that this channel is an unavoidable consequence of the existence of the anomaly-induced  Wess-Zumino-Witten term and compute the relevant cooling rates. We then perform a complementary data-driven study where the rate of photo-production can be estimated from low-energy pion photo-production data. We however conclude that the cooling rates induced by the photo-production of axions and neutrinos are typically sub-dominant with respect to the main channels usually included.}

\FullConference{2nd Training School and General Meeting of the COST Action COSMIC WISPers  (CA21106) (COSMICWISPers2024)\\
 10-14 June 2024 and 3-6 September 2024\\
Ljubljana (Slovenia) and Istanbul (Turkey)\\}

\begin{document}
\maketitle

\section{Introduction}

Compact objects like neutron stars and the core of supernovae are mostly sustained by the Fermi pressure of the degenerate gas of neutrons and are consequently mostly stable on their own. The evolution of the temperature of their surface and interior offers however an opportunity to probe SM as well as BSM physics as such media are exceptionally hot and dense. Exploiting fully this opportunity requires the precise theoretical knowledge of the different cooling channels that can transport the energy from the core to the exterior of the compact object, for example via the emission of light particles like neutrinos or axions and that would help its cooling. 

In this article, we propose to study the cooling produced by the collision of photons with neutrons and producing light particles like neutrinos or axions, the so-called \emph{photo-production} channels. 
Such interactions are unavoidable because of the  Wess-Zumino-Witten (WZW) interactions~\cite{Wess:1971yu,Witten:1983tw,Kaymakcalan:1983qq,Chou:1983qy,Kawai:1984mx,Pak:1984bn,Harvey:2007ca,Harvey:2007rd,Hill:2007zv,Chakraborty:2023wgl} which account for certain processes including anomalies that can not be generated in the usual $\chi$PT. We will focus on two channels of interest: the axion photo-production $\gamma n \to n a$\cite{Chakraborty:2024tyx} and the photo-production of neutrinos $\gamma n \to n \nu \nu$~\cite{Harvey:2007ca, Chakraborty_2023, Bai:2023bbg}, which are presented on Fig.\ref{fig:uncertn2}. 

 \begin{figure}[!h]
\centering
 \includegraphics[scale=0.5]{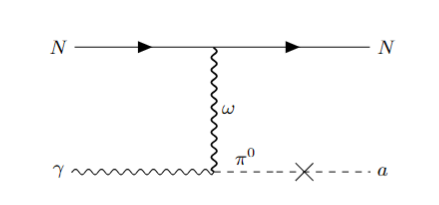}
 \includegraphics[scale=0.5]{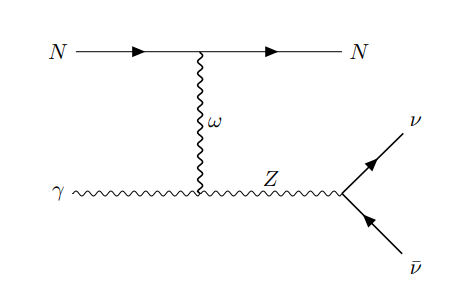}
 \caption{Feynman diagrams of the photo-production of axion (left) and neutrino (right) channels. }
 \label{fig:uncertn2}
\end{figure} 

\section{WZW lagrangian and photo-production channels}
\label{section:photo-production}
We first sketch the derivation of the WZW vertex which we will use in this paper. To generate WZW interactions, one considers the 5-dimensional action~\cite{Wess:1971yu,Witten:1983tw} which is invariant under chiral symmetry, where the boundary is identified with our 4-dimensional spacetime, e.g.,
\begin{eqnarray}
\label{eq:WZW_U2}
 S_{\rm WZW} (A_\mu, U) = N_c \int_D\; d^5 y\; \omega\;, \qquad 
    \omega = -\frac{i}{240\pi^2}\;  \epsilon^{\mu\nu\rho\sigma\tau}\;\text{Tr}\left(\mathcal{U}_\mu\;\mathcal{U}_\nu\;\mathcal{U}_\rho\; \mathcal{U}_\sigma\;\mathcal{U}_\tau\right)\;.
    \label{eq:WZW_U1}
\end{eqnarray}
Here $\mathcal{U}=U^{\dagger}\partial_\mu U$ with $U=\text{Exp}\left(2i\pi^a T^a/f_\pi\right)$. $\pi_a$ are the pion fields with decay constant $f_\pi$ and $A_\mu$'s are the gauge fields. The coefficient $N_c=3$ is fixed by matching with QCD. Any arbitrary subgroup of the chiral symmetry can be gauged using a trial and error method~\cite{Witten:1983tw, Wess:1971yu,Kaymakcalan:1983qq,Chou:1983qy,Kawai:1984mx,Pak:1984bn}. Meson fields such as $\omega$ can be introduced as a background vector field following the method presented in Ref.~\cite{Harvey:2007ca} by replacing $\tilde{A_\mu}=A_\mu+B_\mu$ in the effective action. Proper counter-terms need to be added to maintain gauge invariance and conservation of vector currents. Finally (see \cite{Chakraborty_2023, Chakraborty:2024tyx} for details), we find the following interaction between the pion, the photon and $\omega$,
\begin{align}
    \hspace*{-0.24cm}
         \mathcal{L}_{WZW}^{\pi_0} & \supset \; \frac{N_c\; \epsilon^{\mu\nu\rho\sigma}}{24\pi^2} F_{\mu\nu}\left[\frac{e^2}{4}\frac{\pi_0}{f_\pi}\;F_{\rho\sigma}+eg_\omega\frac{\partial_\rho \pi_0}{f_\pi}\; \omega_\sigma \right]\;.
         \label{eq:pion_WZW}
    \end{align}
    and a similar anomalous interaction between the photon, the Z gauge boson and the  $\omega$\cite{Chakraborty_2023}

\begin{equation}
    \mathcal L_{\text{WZW}} \supset \frac{N_C}{48\pi^2} g_2^2 g_\omega\;\tan\theta_W\;\epsilon^{\mu\nu\rho\sigma}\; F_{\mu\nu}\omega_\rho Z_\sigma\; + ...
    \label{eq:WZW_Lag}
\end{equation}
    
Notice that, similar interactions with other vector mesons are also plausible but suppressed~\cite{Harvey:2007ca}. Nucleons interact with the vector meson fields $\omega$ 
via
$\mathcal{L}_0 = \bar{N}\left(i\gamma_\mu{\partial^\mu}-g_\omega\;\gamma_\mu{\omega^\mu}-m_N\right) N$. 
We will now connect those raw to the axions in section \ref{sec:emission_axion} and to neutrinos in section \ref{sec:emi_neutrino}.

\section{Photo-Production of axions}
\label{sec:emission_axion}

Among all the possible candidates for light degrees of freedom beyond the SM, axions or axion-like particles are the most well-motivated. The axion emerges as the pseudo-Nambu-Goldstone boson of a spontaneously broken $U(1)$ symmetry~\cite{Peccei:1977hh,Peccei:1977ur,Weinberg:1977ma,Wilczek:1977pj} and offers a compelling solution to the long-standing strong-$CP$ problem~\cite{tHooft:1976rip} of the Standard Model (SM). Moreover, axions can also be a natural dark matter candidate~\cite{Preskill:1982cy,Dine:1982ah,Abbott:1982af}, address the hierarchy problem~\cite{Graham:2015cka, Hook:2016mqo, Trifinopoulos:2022tfx} and play a crucial role in resolving the matter-antimatter asymmetry~\cite{Co:2019wyp,Chakraborty:2021fkp} of the Universe. 
Compact objects provide a very interesting mean to probe axions: Nuclear reactions inside the stellar interior such as white dwarfs, neutron stars, supernovae, etc., are potentially powerful sources of axions, which might result in a more efficient transport of energy than the SM neutrinos, and induce observational changes, namely in the spectrum of neutrinos emitted during the first 10 seconds of supernovae. This lead the authors of Ref.~\cite{1990PhR...198....1R,Raffelt:1996wa} to propose the famous bound on the emissivity of axions $
    Q_a/\rho \lesssim 10^{19}\; \text{erg}\; s^{-1} g^{-1} $.
 The main channels of cooling contributing to $Q_a$ are listed in Table \ref{tab_cross_sec}.

\begin{table}[h!]
\begin{center}
\begin{tabular}{lll}
\hline
\multicolumn{1}{|l|}{Process} & \multicolumn{1}{l|}{Coupling} & \multicolumn{1}{l|}{Refs}  \\ \hline
\multicolumn{1}{|l|}{$NN \to NNa$} & \multicolumn{1}{l|}{$\left(C_{aN}/2f_a\right)\; \bar N \gamma_5 \gamma_\mu N\; \partial^\mu a$} & \multicolumn{1}{l|}{\cite{PhysRevLett.53.1198, Iwamoto:1992jp}} 
\\ \hline
\multicolumn{1}{|l|}{$\pi^- p \to Na$} & \multicolumn{1}{l|}{$\left(C_{aN}/2f_a\right)\; \bar N \gamma_5 \gamma_\mu N\; \partial^\mu a$} & \multicolumn{1}{l|}{\cite{PhysRevD.56.2419,PhysRevLett.126.071102}} 
\\ \hline
\multicolumn{1}{|l|}{$N\gamma \to Na$} & \multicolumn{1}{l|}{$i\;\left(C_{aN\gamma}/2\right) a\; \bar N \gamma_5 \sigma_{\mu \nu} N\; F^{\mu \nu}$} & \multicolumn{1}{l|}{\cite{Graham:2013gfa,Lucente:2022vuo}} 
\\ \hline
\multicolumn{1}{|l|}{$N\gamma \to Na$} & \multicolumn{1}{l|} {$\left(\kappa/f_a\right)\epsilon^{\mu \nu \rho \sigma}\; F_{\mu \nu}\;  \partial_\rho a\; \omega_\sigma$+ Data}  & \multicolumn{1}{l|}{This work}
\\ \hline
\end{tabular}
\end{center}
\caption{Interactions relevant for axion emission from a supernova. The coefficient $\kappa$ is defined in Eq.~\eqref{eq:axion_WZW}.
\label{tab_cross_sec}} 
\end{table}

We will now study the photo-production of axions, starting from Eq.\eqref{eq:pion_WZW}. A popular method to incorporate the axion in a amplitude matrix is to take advantage of the mixing between the axion and the pion $\pi_0$,  
$\pi = \pi_{\rm phys} - \theta_{\pi^0-a} a_{\rm phys}$ (see \cite{Aloni:2019ruo}).  The mixing for the QCD axion is\cite{KRAUSS1986483,Notari:2022ffe}
\begin{align}
\theta_{\pi^0-a} &\simeq \frac{1}{2}\frac{f_\pi}{f_a}\frac{m_d-m_u}{m_d+m_u}\; \equiv 
\frac{C^{\rm QCDa}_Af_\pi}{f_a} \, ,
\label{eq:mixing}
\end{align}
while for a massive ALP it is 
\cite{Bauer:2021wjo,DiLuzio:2020wdo,GrillidiCortona:2015jxo, DiLuzio:2024jip}
\begin{align}
\theta_{\pi^0-a} &\simeq \frac{1}{2}\frac{f_\pi}{f_a}\frac{m_a^2}{m_\pi^2-m_a^2}\frac{m_d-m_u}{m_d+m_u}\; \equiv 
\frac{C^{\rm ALP}_Af_\pi}{f_a} \, ,
\label{eq:mixing}
\end{align}
where we made the standard choice of rotating away the mass mixing term. Moreover, $f_\pi \approx 93$ MeV is the pion decay constant. We then trade the mixing angle between axion and pion fields, which generates axion-photon coupling and the desired interactions of the form
\begin{equation}
\mathcal L_{\rm WZW}^a \supset \frac{\kappa}{f_a}\; \epsilon^{\mu\nu\rho\sigma}\; F_{\mu\nu}\;\partial_\rho a \; \omega_\sigma\; , \; \kappa = \frac{C_A N_c}{24\pi^2}\; e g_\omega\; .  
    \label{eq:axion_WZW}
\end{equation}
We however notice that this method is intrinsically ambiguous as it contains a leftover dependence on a non-physical parameter in the chiral rotation. This issue has been solved for the axion-photon coupling in \cite{Bauer:2017ris, Bauer:2021wjo} and a similar solution for the $\omega-\gamma-a$ coupling as been proposed in \cite{Bai:2024lpq}. 
 
Thus the emissivity from photo-production of axions in the non-degenerate regime finally simplifies to
\begin{align}
 Q_{N\gamma \to N a} \approx   \int \frac{ d E_\gamma }{\pi^2 } E_\gamma^3 f_\gamma (p_\gamma) n_B \sigma_{\gamma N \to N a} (E_\gamma) \;,  \qquad  n_B \equiv \int  \frac{d^3 p_{N}}{(2\pi)^3  }  g_N  f_N \, .
 \label{eq:emissivity_exp}
\end{align}
and we obtain after computation the cooling rate 
\begin{align}
    Q_{N \gamma \to N a}^{\rm WZW, ND} \approx   \; g_\gamma g_\omega^2\;  \frac{\kappa^2 n_b}{2\pi^3}\; \frac{T^2}{f_a^2}\; \frac{m_a^4}{m_\omega^4}\; \left[4m_a T\; K_3\left(\frac{m_a}{T}\right)  +(m_a^2+35T^2)\;K_4\left(\frac{m_a}{T}\right)\right]\;,
    \label{eq:ND}
\end{align}
where $K_{3,4}$ are the Bessel function of type three and four respectively. It was recently confirmed in \cite{Cao:2024cym} that the WZW anomaly induced coupling $ \gamma-\omega-a$ dominates the rate of the photo-production $n\gamma \to n a$ in most of the regime of energies of interest.

\begin{figure}[!h]
\hspace{-0.9cm} \includegraphics[scale=0.5]{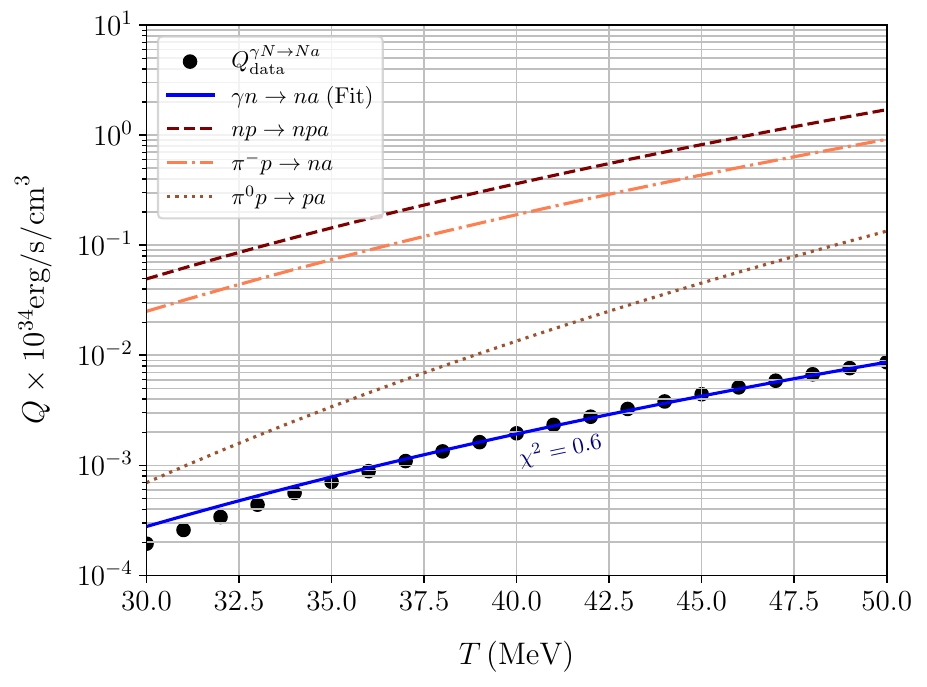}
\includegraphics[scale=0.5]{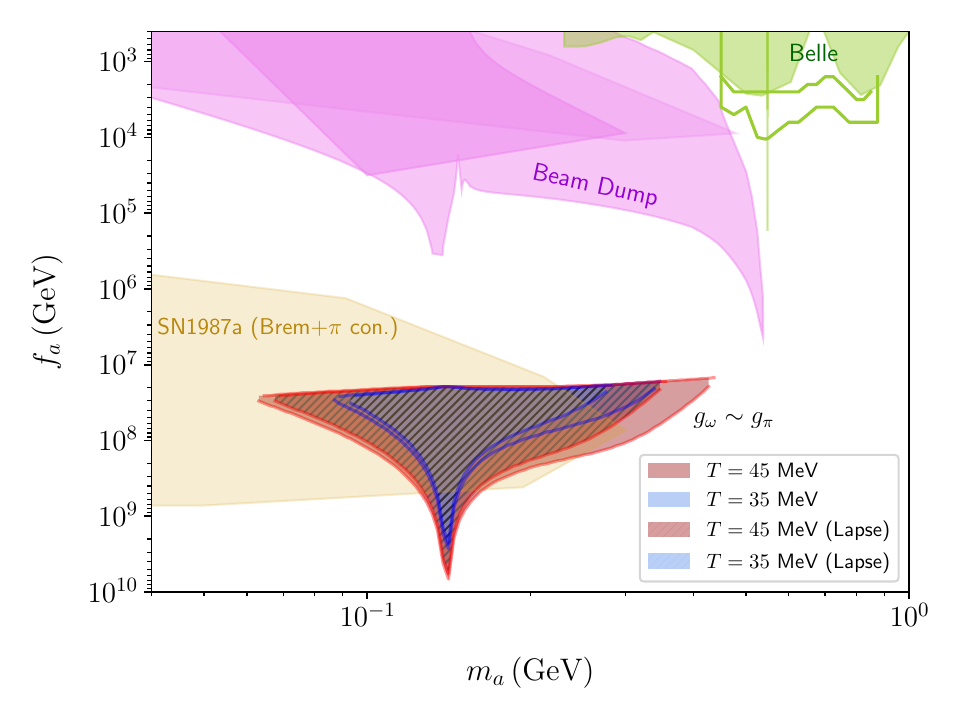}
 \caption{Left: comparison of the different cooling rates as a function of the temperature. Right: exclusion region from the photo-production alone.}
 \label{fig:axion_emission}
\end{figure}

However as a complementary analysis, we can use the data collected on the photo-production of $\pi_0$ in energy range $E_\gamma=145-180$ MeV~\cite{Briscoe:2020qat} and in the range  $E_\gamma=180-500$~\cite{PhysRevC.100.065205} to fit the cross-section and to obtain am estimate of the cross-section of axion photo-production using $\sigma_{n\gamma \to n a} = \theta_{a-\pi_0}^2 \sigma_{n\gamma \to n \pi_0} $. This leads to the following cooling rate
\begin{align}
\label{eq:fit_data}
\frac{Q_{\rm data} }{  10^{33} \text{ cm}^{-3}s^{-1}\text{erg}} \approx   1.54 \bigg(\frac{C_A  10^9}{f_a/\text{GeV}} \bigg)^2  \rho_{15}\;  T_{40}^{6.73} \,, 
\end{align}
 It appears that this data-driven contribution to axion emissivity dominates over the WZW term. This cooling rate can have application for the bound on axions from supernovae and we compare photo-production cooling rates with the rates from other processes in supernovae in fig.\ref{fig:axion_emission}. 

\section{Photo-Production of neutrinos}
\label{sec:emi_neutrino}
As a side of the photo-production of axion, one also can envision the photo-production of neutrinos. Starting from Eq.\eqref{eq:WZW_Lag}, which allows us to integrate out the heavy vector meson $\omega$ and the $Z$, we obtain the effective coupling
\begin{equation}
    \mathcal L_{\rm int} = \bigg(\frac{N_C}{48\pi^2} g_2^2 \frac{g_\omega^2}{m_\omega^2M_Z^2}\;\tan\theta_W \bigg) \times \;\epsilon^{\mu\nu\rho\sigma}\; F_{\mu\nu}\bar N \gamma_\rho N\bar \nu \gamma_\sigma  \nu\, . 
\end{equation}

Then the cooling rate in the regime where the neutrons are degenerate can be computed \cite{Chakraborty_2023}
\begin{equation}
  Q^{\gamma n \to n \nu \nu} =
    \begin{cases}
      1.2\times 10^{5} \,\dfrac{48\,n_F \, M_N^2 \, \kappa^2}{(2\,\pi)^7}\,T^{11} & \text{for $m_\gamma/T \lesssim 5$}\;,\\
      \dfrac{48\,n_F \, M_N^2 \, \kappa^2}{(2\,\pi)^7}\,\sqrt{\dfrac{\pi}{2}}\,T^{7/2}\,m_\gamma^{15/2}\,e^{-m_\gamma/T}& \text{for $m_\gamma/T \gg 5$}\;.\\
    \end{cases}  
    \label{eq:Qfinal}
\end{equation}
 This cooling rate might have application for the cooling of young neutron stars with MeV temperatures as we show on Fig.\ref{fig:uncertn2} in blue bands. One displays the rates
 of cooling from photo-production against the other relevant channels (mURCA\cite{Yakovlev:2000jp}, PBF\cite{Voskresensky_1998}, and bremsstrahlung\cite{Yakovlev:2000jp,1979ApJ...232..541F,1995A&A...297..717Y}). We emphasize that this rate of cooling is suppressed with respect to the rate found in \cite{Harvey:2007rd} (pink band) by several orders of magnitude. This discrepancy originates from the degeneracy suppression in NS.
 \begin{figure}[!h]
\centering
 \includegraphics[scale=0.5]{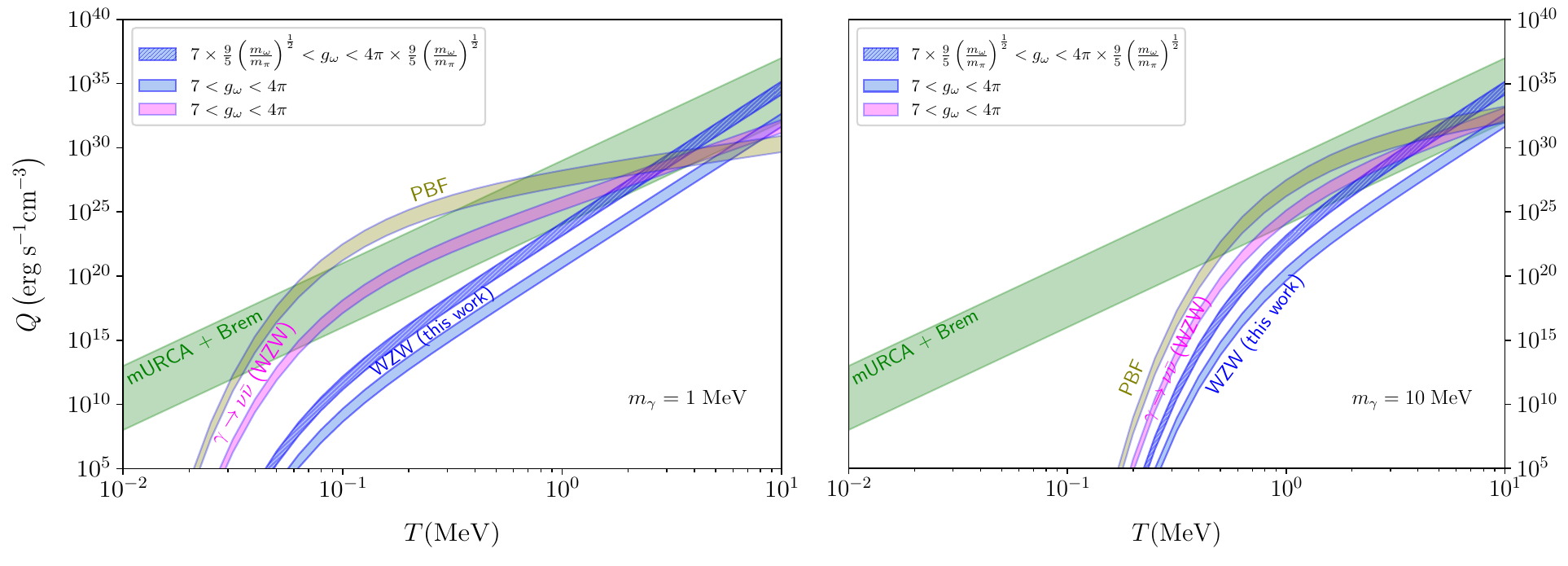}
 \caption{Comparison of emissivities due to the WZW term (blue) with other conventional processes like mURCA~\cite{Yakovlev:2000jp}, PBF~\cite{Voskresensky_1998}, and $\nu$ bremsstrahlung~\cite{Yakovlev:2000jp,1979ApJ...232..541F,1995A&A...297..717Y} for a 1 MeV photon (left) and for a 10 MeV photon (right). The pink band $\gamma\to\nu\bar\nu$ (WZW) shows the (wrong) result of \cite{Harvey:2007rd}. }
 \label{fig:uncertn2}
\end{figure}

\section{Summary and outlook} 

Going back to Fig.\ref{fig:uncertn2}, one observe that photo-production is likely to be a subleading process in the cooling of young NS, as opposed to a claim made in \cite{Harvey:2007rd}. The reason of the difference is that \cite{Harvey:2007rd} overlooked degeneracy suppression. Fig.\ref{fig:axion_emission} implies similar conclusions for the emission of axions in the supernovae, where it is likely that axion-photo-production is subdominant against bremsstrahlung. 

The gauged WZW contains however a zoo of interactions which has been only explored in a limited amount of phenomenological works\cite{Harvey:2007ca, Harvey:2007rd, Chakraborty_2023, Chakraborty:2024tyx, Bai:2023bbg}, and certainly deserves further exploration.

\bibliographystyle{JHEP}
{\footnotesize
\bibliography{biblio}}

\providecommand{\href}[2]{#2}\begingroup\raggedright\begin{thebibliography}{10}

\bibitem{Wess:1971yu}
J.~Wess and B.~Zumino {\em Phys. Lett. B} {\bf 37} (1971) 95--97.

\bibitem{Witten:1983tw}
E.~Witten {\em Nucl. Phys. B} {\bf 223} (1983) 422--432.

\bibitem{Kaymakcalan:1983qq}
O.~Kaymakcalan, S.~Rajeev, and J.~Schechter {\em Phys. Rev. D} {\bf 30} (1984)
  594.

\bibitem{Chou:1983qy}
K.-c. Chou, H.-y. Guo, K.~Wu, and X.-c. Song {\em Phys. Lett. B} {\bf 134}
  (1984) 67--69.

\bibitem{Kawai:1984mx}
H.~Kawai and S.~H.~H. Tye {\em Phys. Lett. B} {\bf 140} (1984) 403--407.

\bibitem{Pak:1984bn}
N.~K. Pak and P.~Rossi {\em Nucl. Phys. B} {\bf 250} (1985) 279--294.

\bibitem{Harvey:2007ca}
J.~A. Harvey, C.~T. Hill, and R.~J. Hill {\em Phys. Rev. D} {\bf 77} (2008)
  085017, [\href{http://arxiv.org/abs/0712.1230}{{\tt arXiv:0712.1230}}].

\bibitem{Harvey:2007rd}
J.~A. Harvey, C.~T. Hill, and R.~J. Hill {\em Phys. Rev. Lett.} {\bf 99} (2007)
  261601, [\href{http://arxiv.org/abs/0708.1281}{{\tt arXiv:0708.1281}}].

\bibitem{Hill:2007zv}
C.~T. Hill and R.~J. Hill {\em Phys. Rev. D} {\bf 76} (2007) 115014,
  [\href{http://arxiv.org/abs/0705.0697}{{\tt arXiv:0705.0697}}].

\bibitem{Chakraborty:2023wgl}
S.~Chakraborty, A.~Gupta, and M.~Vanvlasselaer {\em JCAP} {\bf 10} (2023) 030,
  [\href{http://arxiv.org/abs/2306.15872}{{\tt arXiv:2306.15872}}].

\bibitem{Chakraborty:2024tyx}
S.~Chakraborty, A.~Gupta, and M.~Vanvlasselaer
  \href{http://arxiv.org/abs/2403.12169}{{\tt arXiv:2403.12169}}.

\bibitem{Chakraborty_2023}
S.~Chakraborty, A.~Gupta, and M.~Vanvlasselaer {\em Journal of Cosmology and
  Astroparticle Physics} {\bf 2023} (oct, 2023) 030.

\bibitem{Bai:2023bbg}
Y.~Bai and C.~H. de~Lima \href{http://arxiv.org/abs/2311.18794}{{\tt
  arXiv:2311.18794}}.

\bibitem{Peccei:1977hh}
R.~D. Peccei and H.~R. Quinn {\em Phys. Rev. Lett.} {\bf 38} (1977) 1440--1443.

\bibitem{Peccei:1977ur}
R.~D. Peccei and H.~R. Quinn {\em Phys. Rev. D} {\bf 16} (1977) 1791--1797.

\bibitem{Weinberg:1977ma}
S.~Weinberg {\em Phys. Rev. Lett.} {\bf 40} (1978) 223--226.

\bibitem{Wilczek:1977pj}
F.~Wilczek {\em Phys. Rev. Lett.} {\bf 40} (1978) 279--282.

\bibitem{tHooft:1976rip}
G.~'t~Hooft {\em Phys. Rev. Lett.} {\bf 37} (1976) 8--11.

\bibitem{Preskill:1982cy}
J.~Preskill, M.~B. Wise, and F.~Wilczek {\em Phys. Lett. B} {\bf 120} (1983)
  127--132.

\bibitem{Dine:1982ah}
M.~Dine and W.~Fischler {\em Phys. Lett. B} {\bf 120} (1983) 137--141.

\bibitem{Abbott:1982af}
L.~F. Abbott and P.~Sikivie {\em Phys. Lett. B} {\bf 120} (1983) 133--136.

\bibitem{Graham:2015cka}
P.~W. Graham, D.~E. Kaplan, and S.~Rajendran {\em Phys. Rev. Lett.} {\bf 115}
  (2015), no.~22 221801, [\href{http://arxiv.org/abs/1504.07551}{{\tt
  arXiv:1504.07551}}].

\bibitem{Hook:2016mqo}
A.~Hook and G.~Marques-Tavares {\em JHEP} {\bf 12} (2016) 101,
  [\href{http://arxiv.org/abs/1607.01786}{{\tt arXiv:1607.01786}}].

\bibitem{Trifinopoulos:2022tfx}
S.~Trifinopoulos and M.~Vanvlasselaer {\em Phys. Rev. D} {\bf 107} (2023),
  no.~7 L071701, [\href{http://arxiv.org/abs/2210.13484}{{\tt
  arXiv:2210.13484}}].

\bibitem{Co:2019wyp}
R.~T. Co and K.~Harigaya {\em Phys. Rev. Lett.} {\bf 124} (2020), no.~11
  111602, [\href{http://arxiv.org/abs/1910.02080}{{\tt arXiv:1910.02080}}].

\bibitem{Chakraborty:2021fkp}
S.~Chakraborty, T.~H. Jung, and T.~Okui {\em Phys. Rev. D} {\bf 105} (2022),
  no.~1 015024, [\href{http://arxiv.org/abs/2108.04293}{{\tt
  arXiv:2108.04293}}].

\bibitem{1990PhR...198....1R}
G.~G. {Raffelt} {\em physrep} {\bf 198} (Dec., 1990) 1--113.

\bibitem{Raffelt:1996wa}
G.~G. Raffelt, {\em {Stars as laboratories for fundamental physics}: {The
  astrophysics of neutrinos, axions, and other weakly interacting particles}}.
\newblock 5, 1996.

\bibitem{PhysRevLett.53.1198}
N.~Iwamoto {\em Phys. Rev. Lett.} {\bf 53} (Sep, 1984) 1198--1201.

\bibitem{Iwamoto:1992jp}
N.~Iwamoto {\em Phys. Rev. D} {\bf 64} (2001) 043002.

\bibitem{PhysRevD.56.2419}
W.~Keil, H.-T. Janka, D.~N. Schramm, G.~Sigl, M.~S. Turner, and J.~Ellis {\em
  Phys. Rev. D} {\bf 56} (Aug, 1997) 2419--2432.

\bibitem{PhysRevLett.126.071102}
P.~Carenza, B.~Fore, M.~Giannotti, A.~Mirizzi, and S.~Reddy {\em Phys. Rev.
  Lett.} {\bf 126} (Feb, 2021) 071102.

\bibitem{Graham:2013gfa}
P.~W. Graham and S.~Rajendran {\em Phys. Rev. D} {\bf 88} (2013) 035023,
  [\href{http://arxiv.org/abs/1306.6088}{{\tt arXiv:1306.6088}}].

\bibitem{Lucente:2022vuo}
G.~Lucente, L.~Mastrototaro, P.~Carenza, L.~Di~Luzio, M.~Giannotti, and
  A.~Mirizzi {\em Phys. Rev. D} {\bf 105} (2022), no.~12 123020,
  [\href{http://arxiv.org/abs/2203.15812}{{\tt arXiv:2203.15812}}].

\bibitem{Aloni:2019ruo}
D.~Aloni, C.~Fanelli, Y.~Soreq, and M.~Williams {\em Phys. Rev. Lett.} {\bf
  123} (2019), no.~7 071801, [\href{http://arxiv.org/abs/1903.03586}{{\tt
  arXiv:1903.03586}}].

\bibitem{KRAUSS1986483}
L.~M. Krauss and M.~B. Wise {\em Physics Letters B} {\bf 176} (1986), no.~3
  483--485.

\bibitem{Notari:2022ffe}
A.~Notari, F.~Rompineve, and G.~Villadoro {\em Phys. Rev. Lett.} {\bf 131}
  (2023), no.~1 011004, [\href{http://arxiv.org/abs/2211.03799}{{\tt
  arXiv:2211.03799}}].

\bibitem{Bauer:2021wjo}
M.~Bauer, M.~Neubert, S.~Renner, M.~Schnubel, and A.~Thamm {\em Phys. Rev.
  Lett.} {\bf 127} (2021), no.~8 081803,
  [\href{http://arxiv.org/abs/2102.13112}{{\tt arXiv:2102.13112}}].

\bibitem{DiLuzio:2020wdo}
L.~Di~Luzio, M.~Giannotti, E.~Nardi, and L.~Visinelli {\em Phys. Rept.} {\bf
  870} (2020) 1--117, [\href{http://arxiv.org/abs/2003.01100}{{\tt
  arXiv:2003.01100}}].

\bibitem{GrillidiCortona:2015jxo}
G.~Grilli~di Cortona, E.~Hardy, J.~Pardo~Vega, and G.~Villadoro {\em JHEP} {\bf
  01} (2016) 034, [\href{http://arxiv.org/abs/1511.02867}{{\tt
  arXiv:1511.02867}}].

\bibitem{DiLuzio:2024jip}
L.~Di~Luzio, A.~W.~M. Guerrera, X.~Ponce~D\'\i{}az, and S.~Rigolin
  \href{http://arxiv.org/abs/2402.12454}{{\tt arXiv:2402.12454}}.

\bibitem{Bauer:2017ris}
M.~Bauer, M.~Neubert, and A.~Thamm {\em JHEP} {\bf 12} (2017) 044,
  [\href{http://arxiv.org/abs/1708.00443}{{\tt arXiv:1708.00443}}].

\bibitem{Bai:2024lpq}
Y.~Bai, T.-K. Chen, J.~Liu, and X.~Ma
  \href{http://arxiv.org/abs/2406.11948}{{\tt arXiv:2406.11948}}.

\bibitem{Cao:2024cym}
X.-H. Cao and Z.-H. Guo \href{http://arxiv.org/abs/2408.15825}{{\tt
  arXiv:2408.15825}}.

\bibitem{Briscoe:2020qat}
W.~J. Briscoe, A.~E. Kudryavtsev, I.~I. Strakovsky, V.~E. Tarasov, and R.~L.
  Workman {\em Eur. Phys. J. A} {\bf 56} (2020), no.~8 218,
  [\href{http://arxiv.org/abs/2004.01742}{{\tt arXiv:2004.01742}}].

\bibitem{PhysRevC.100.065205}
{\bf A2 Collaboration at MAMI} Collaboration, W.~J. Briscoe, M.~Had\ifmmode
  \check{z}\else \v{z}\fi{}imehmedovi\ifmmode~\acute{c}\else \'{c}\fi{}, A.~E.
  Kudryavtsev, V.~V. Kulikov, M.~A. Martemianov, I.~I. Strakovsky,
  A.~\ifmmode~\check{S}\else \v{S}\fi{}varc, V.~E. Tarasov, R.~L. Workman,
  S.~Abt, P.~Achenbach, C.~S. Akondi, F.~Afzal, P.~Aguar-Bartolom\'e, Z.~Ahmed,
  J.~R.~M. Annand, H.~J. Arends, K.~Bantawa, M.~Bashkanov, R.~Beck, M.~Biroth,
  N.~Borisov, A.~Braghieri, S.~A. Bulychjov, F.~Cividini, C.~Collicott,
  S.~Costanza, A.~Denig, E.~J. Downie, P.~Drexler, S.~Fegan, M.~I.
  Ferretti~Bondy, S.~Gardner, D.~Ghosal, D.~I. Glazier, I.~Gorodnov, W.~Gradl,
  M.~G\"unther, D.~Gurevich, L.~Heijkenskj\"old, D.~Hornidge, G.~M. Huber,
  A.~K\"aser, V.~L. Kashevarov, S.~Kay, M.~Korolija, B.~Krusche, A.~Lazarev,
  K.~Livingston, S.~Lutterer, I.~J.~D. MacGregor, R.~Macrae, D.~M. Manley,
  P.~P. Martel, J.~C. McGeorge, D.~G. Middleton, R.~Miskimen, E.~Mornacchi,
  A.~Mushkarenkov, C.~Mullen, A.~Neganov, A.~Neiser, M.~Ostrick, P.~B. Otte,
  H.~Osmanovi\ifmmode~\acute{c}\else \'{c}\fi{},
  R.~Omerovi\ifmmode~\acute{c}\else \'{c}\fi{}, B.~Oussena, D.~Paudyal,
  P.~Pedroni, A.~Powell, S.~N. Prakhov, G.~Ron, T.~Rostomyan, A.~Sarty,
  C.~Sfienti, V.~Sokhoyan, K.~Spieker, J.~Stahov, O.~Steffen, I.~Supek,
  A.~Thiel, M.~Thiel, A.~Thomas, L.~Tiator, M.~Unverzagt, Y.~A. Usov, N.~K.
  Walford, D.~P. Watts, S.~Wagner, D.~Werthm\"uller, J.~Wettig, M.~Wolfes, and
  N.~Zachariou {\em Phys. Rev. C} {\bf 100} (Dec, 2019) 065205.

\bibitem{Yakovlev:2000jp}
D.~G. Yakovlev, A.~D. Kaminker, O.~Y. Gnedin, and P.~Haensel {\em Phys. Rept.}
  {\bf 354} (2001) 1, [\href{http://arxiv.org/abs/astro-ph/0012122}{{\tt
  astro-ph/0012122}}].

\bibitem{Voskresensky_1998}
D.~N. Voskresensky, E.~E. Kolomeitsev, and B.~Kämpfer {\em Journal of
  Experimental and Theoretical Physics} {\bf 87} (aug, 1998) 211--217.

\bibitem{1979ApJ...232..541F}
B.~L. {Friman} and O.~V. {Maxwell} {\em apj} {\bf 232} (Sept., 1979) 541--557.

\bibitem{1995A&A...297..717Y}
D.~G. {Yakovlev} and K.~P. {Levenfish} {\em aap} {\bf 297} (May, 1995) 717.

\end{thebibliography}\endgroup

\end{document}